\documentclass[11pt]{article}

\usepackage{amsmath, euscript, amssymb, amsfonts}

\usepackage[matrix,arrow,curve]{xy}
\usepackage{mathrsfs}



\newcommand{\supp}{\mathop{\rm supp}\nolimits}

\newcommand{\mS}{{\mathscr{S}}}

\newcommand{\oR}{{\mathbb R}}
\newcommand{\oV}{{\mathbb V}}
\newcommand{\oC}{{\mathbb C}}
\newcommand{\oZ}{{\mathbb Z}}

\newcommand{\im}{\mathop{\mathrm{im}}\nolimits}
\newcommand{\eqdef}{\stackrel{\mathrm{def}}{=}}

\begin{document}

\bigskip

\phantom{X}

\thispagestyle{empty}

\baselineskip=15pt

\vspace{2cm}

\begin{center}
{\Large\bf  Noncommutative deformations of quantum field theories,
locality, and causality}

\vspace{1cm}

{\large\bf M.~A.~Soloviev}\footnote{E-mail: soloviev@lpi.ru}

\vspace{0.5cm}

 \centerline{\sl P.~N.~Lebedev Physical Institute}
 \centerline{\sl Russian Academy of Sciences}
 \centerline{\sl  Leninsky Prospect 53, Moscow 119991, Russia}

\vskip 3em

\end{center}

\begin{abstract}
We investigate noncommutative deformations of quantum field
theories for different star products,  particularly emphasizing
the locality properties and the regularity  of the deformed
fields. Using functional analysis methods, we  describe  the basic
structural features of the vacuum expectation values of
star-modified products of fields and field commutators. As an
alternative to microcausality, we introduce a notion of
$\theta$-locality, where $\theta$ is the noncommutativity
parameter. We also analyze the conditions for the convergence and
continuity of star products  and define the function algebra that
is most suitable for the Moyal and Wick-Voros products. This
algebra corresponds to the concept of strict deformation
quantization and is a useful tool for constructing quantum field
theories on a noncommutative space-time.
\end{abstract}

\vspace{1cm}


\newpage

\setcounter{page}{2}

\section{Introduction}
\label{sec1} The recent intensive study  of noncommutative field
theory models have been motivated by fundamental issues in quantum
physics and gravity~\cite{1}. Although the idea of space-time
noncommutativity is old and, as well known, goes back to
Heisenberg and Snyder, the development of string theory has
inspired a renewed interest in this  research field (see, e.g.,
\cite{2} for a review). Almost  all works on this subject begin by
replacing the ordinary pointwise product of functions on
space-time with a noncommutative star product. As a result, the
coordinate functions satisfy the commutation relation
\begin{equation}
[x^\mu,  x^\nu]_\star\equiv x^\mu\star  x^\nu-x^\nu\star x^\mu
=i\theta^{\mu\nu},
 \label{1}
\end{equation}
where $\theta^{\mu\nu}$ is a real antisymmetric matrix  playing
the role of a noncommutativity parameter and determining a
deformation of the Minkowski space. But this deformation can be
combined with the basic principles of quantum theory many ways.
Moreover, there are many star products compatible with commutation
relation~\cite{1}. The great majority of authors use the
Moyal-Weyl-Gr\"onewold star product, but a field theory with the
Wick-Voros (or normally ordered) star product was  recently
discussed~\cite{3,4}. Interestingly, the authors reached opposite
conclusions about the equivalence of theories with different star
products.

There is not yet a consensus  about the best way of constructing a
self-consistent noncommutative field theory.  The  so-called
twisted Poincar\'e covariance, aimed at restoring the space-time
symmetries broken by noncommutativity, is much considered
(see~\cite{5,6} and the references therein). How to implement this
covariance is also  debated, but  the notion of a twisted tensor
product associated with a given star product plays a major role in
all implementations. Since we lack  a generally accepted
formulation of noncommutative quantum field theory, the question
of the causality of observables merits further investigation,
because it is crucial for the physical interpretation.

We note that the star  products and twisted tensor products can be
treated from two different standpoints. It is more common to
regard them as formal series in powers of the deformation
parameter $\theta$. But this approach is  not quite acceptable
physically. When analyzing the causal structure of noncommutative
models, we should  take into account that these products are
inherently nonlocal because they are determined by infinite-order
differential operators. To gain a better insight into these
nonlocal features, we must establish the conditions under which
the formal series  converge and  describe the corresponding
topological algebras in which  the star product  depends
continuously  on the deformation parameter. In what follows, we
adhere to the second approach, which is close to the strategy of
strict deformation quantization~\cite{7}.

The plan of presentation is as follows. We first  recall the
definitions of the Moyal and Wick-Voros star products and their
associated tensor products. We then show that the Moyal twisted
tensor product can be used to naturally deform~\cite{8,9} a
quantum field theory  initially defined on commutative Minkowski
space and describe basic properties of the resulting
noncommutative field theory. In particular, the deformed quantum
fields can be localized in certain wedge-shaped regions~\cite{9},
but this localization differs essentially  from the modification
of the light cone to a light wedge that was previously
proposed~\cite{10,11,12} for theories  with space-space
noncommutativity describing the low-energy limit of string theory.
Further, we consider the convergence conditions for the formal
power series determining star products and describe the basic
properties of the corresponding function algebras~\cite{13}. This
analysis and some instructive examples lead  to the concept of
$\theta$-locality~\cite{14} which in our opinion can be used to
implement causality in a noncommutative quantum field theory.

\section{The Moyal and Wick-Voros star products}
\label{sec2}

The formal power series representing the Moyal and Wick-Voros star
products  are written  as
\begin{multline}
(f\star_M
g)(x)=f(x)\exp\left(\frac{i}{2}\,\overleftarrow{\partial_\mu}\,
\theta^{\mu\nu}\,\overrightarrow{\partial_\nu}\right)g(x)\\
=f(x)g(x)+\sum_{n=1}^\infty\left(\frac{i}{2}\right)^n\frac{1}{n!}\,
\theta^{\mu_1\nu_1}\dots \theta^{\mu_n\nu_n}\partial_{\mu_1}\dots
\partial_{\mu_n}f(x)\partial_{\nu_1}\dots\partial_{\nu_n}g(x)
\label{2}
\end{multline}
and
\begin{equation}
(f\star_V
g)(x)=f(x)\exp\left(\frac{i}{2}\,\overleftarrow{\partial_\mu}\,
\theta^{\mu\nu}\,\overrightarrow{\partial_\nu}+\frac{\vartheta}{2}\,
\overleftarrow{\partial_\mu}\,
\delta^{\mu\nu}\overrightarrow{\partial_\nu}\right)g(x). \label{3}
\end{equation}
  For simplicity in the second case, we assume that the matrix
$\theta^{\mu\nu}$ has a canonical block-diagonal form with the
same parameter $\vartheta>0$ in every $2\times2$ block
$\begin{pmatrix} 0&\vartheta \\ -\vartheta&0 \end{pmatrix}$,
although  different blocks may have different values of
$\vartheta$, including   zero in applications. These two star
products play a central role in the Weyl-Wigner approach to
quantum mechanics because the Moyal product is compatible with the
symplectic structure of the linear phase space and both of them
are compatible with its complex structure. It is easy to see that
these two products yield the same commutation relation~\eqref{1}
for the coordinate functions. There is a simple relation between
these products:
 \begin{equation}
T(f\star_M g)=T(f)\star_V T(g),\qquad \text{where}\quad T=
e^{\tfrac{\vartheta}{4}\Delta},
\quad\Delta=\sum_{\mu=1}^d\partial_\mu^2.
 \label{4}
 \end{equation}
This relation  holds in the sense of formal power series and was
discussed, in particular, in~\cite{15} in the quantum mechanics
context. But from the standpoint of strict deformation
quantization,  products \eqref{2} and~\eqref{3}   and the operator
$T$ are well defined only under rather severe restrictions on the
functions. In the quantum field theory formalism, the standard
practice is  to use the Schwartz space $S(\oR^d)$ of rapidly
decreasing smooth functions and it is commonly assumed that this
space is an algebra under the Moyal multiplication. But we note
that  series~\eqref{2}  generally diverges for functions in
$S(\oR^d)$. In fact, using the classical Borel theorem (see,
e.g.,~\cite{16}) we can construct functions  $f, g\in S(\oR^d)$
such that the sequence $f( \overleftarrow{\partial_\mu}\,
\theta^{\mu\nu}\,\overrightarrow{\partial_\nu})^ng|_{x=x_0}$,
$n=0,1,\dots$, coincides with any given number sequence. Moreover,
it is easily shown~\cite{13} that  the series defining $f\star_Mf$
does not converge in the topology of $S(\oR^d)$ even for the
Gaussian function $f(x)=e^{-\gamma|x|^2}$ if $\gamma
>1/\vartheta$. Nevertheless the multiplication $\star_M$ is well defined
on an analytic function  space that we describe in Sec.~\ref{sec6}
This space is dense in $S(\oR^d)$ and the multiplication $\star_M$
has a unique continuous extension to the  Schwartz space; this
extension  can be written in terms of the Fourier transforms as
\begin{equation}
(f\star_M g)(x)=\frac{1}{(2\pi)^{2d}}\iint\hat f(k)\hat g(k')
e^{i(k+k')\cdot x-\tfrac{i}{2}k\theta k'} \,dk\,dk', \label{5}
\end{equation}
where $k\theta k'\eqdef k_\mu\theta^{\mu\nu}k'_\nu$ is a
symplectic inner product on $\oR^d$.

It is easily seen that $S(\oR^d)$ is an associative topological
algebra with respect to the product defined by~\eqref{5}. The
operator $T=e^{\vartheta\Delta/4}$ also has a natural extension to
$S(\oR^d)$ coinciding with multiplication of the Fourier
transforms by $e^{-\vartheta |k|^2/4}$. Its image $\im T$ consists
of all entire functions satisfying the inequalities
\begin{equation*}
|f(x+iy)|\leq C_N(1+|x|)^{-N}e^{|y|^2/\vartheta},\qquad
N=0,1,\dots.
\end{equation*}
Such functions form an algebra under the Wick-Voros product or,
more precisely, under the multiplication
\begin{equation}
(f\star_V g)(x)=\frac{1}{(2\pi)^{2d}}\iint\hat f(k)\hat g(k')
e^{i(k+k')\cdot x-\tfrac{i}{2}k\theta k'-\tfrac{\vartheta}{2}k k'}
\,dk\,dk', \label{6}
\end{equation}
where $k k'= \sum_\mu k_\mu k'_\mu$ is the Euclidean scalar
product on  $\oR^d$. But the whole space $S(\oR^d)$ cannot be made
into a $\star_V$-algebra.

{\bf Proposition.} \label{prop1} {\it The Wick-Voros star product
does not admit a continuous extension to the Schwartz space.

Proof.} Let $u$ be the linear functional $f\to \int f(x)\,dx=\hat
f(0)$. Clearly,  $u\in S'$. For any $f\in \im T$, we have
\begin{equation*}
u(f\star_V f)=\frac{1}{(2\pi)^{d}}\int\hat f(k)\hat f(-k)\,
e^{\tfrac{\vartheta}{2}|k|^2} \,dk.
\end{equation*}
We set $\hat f(k)=e^{-\vartheta |k|^2/4}$ and consider the
sequence $\hat f_n(k)=e^{-|k|^4/n}\hat f(k)$ which obviously
belongs to $\im T$ and converges to $\hat f$ in the
Fourier-invariant space $S(\oR^d)$. We suppose that there is  a
continuous extension of product~\eqref{6} to this space. Then
there would exist a function $g\in S(\oR^d)$ such that $g=f\star_V
f$ and $u(f_n\star_V f_n)\to \hat g(0)$. But
\begin{equation*}
u(f_n\star_V f_n)=\frac{1}{(2\pi)^{d}}\int
e^{-\tfrac{2}{n}|k|^4}\,dk\to\infty,
\end{equation*}
and this contradiction completes the proof.

Similar remarks can be made about the twisted tensor product. In
the Moyal case it is defined by
\begin{equation}
(f_1\otimes_M\cdots\otimes_M f_n)(x_1,\dots,x_n)=\prod_{1\leq
a<b\leq n}
e^{\tfrac{i}{2}\partial_{x_a}\theta\partial_{x_b}}f_1(x_1)\cdots
f_n(x_n),
\label{7}
\end{equation}
where
\begin{equation*}
\partial_{x_a}\theta\partial_{x_b}=
\frac{\partial}{\partial
x_a^\mu}\,\theta^{\mu\nu}\frac{\partial}{\partial x_b^\nu},
\end{equation*}
and is often called the ``star product at different points'' The
star product itself is obtained from the twisted tensor product by
identifying  these points,
\begin{equation*}
(f_1\star_M\dots\star_M f_n)(x) = (f_1\otimes_M\dots\otimes_M
f_n)|_{x_1=\dots= x_n=x},
\end{equation*}
in complete analogy with the case of the ordinary pointwise
product, which is obtained from the ordinary tensor product of
functions by the same identification. In the Wick-Voros case we
have
\begin{equation}
(f_1\otimes_V\cdots\otimes_V f_n)(x_1,\dots,x_n)=\prod_{1\leq
a<b\leq n} e^{\tfrac{i}{2}\partial_{x_a}\theta\partial_{x_b}+
\tfrac{\vartheta}{2}\partial_a\partial_b}f_1(x_1)\cdots f_n(x_n),
\label{8}
\end{equation}
where
\begin{equation*}
\partial_a\partial_b=\sum_\mu\frac{\partial}{\partial
x_a^\mu}\frac{\partial}{\partial x_b^\mu}.
\end{equation*} The twisted tensor product is
associative and can be generalized to functions of several
variables:
\begin{equation}
(f\otimes_M
g)(x_1,\dots,x_m;x'_1,\dots,x'_n)=\prod_{a=1}^m\prod_{b=1}^n
e^{\tfrac{i}{2}\partial_{x_a}\theta\partial_{x'_b}}
f(x_1,\dots,x_m)g(x'_1,\dots,x'_n),
 \label{9}
\end{equation}
where the right-hand side is determined by the requirement of
compatibility with~\eqref{7} for $f=f_1\otimes_M\dots\otimes_M
f_m$ and $g=g_1\otimes_M\dots\otimes_M g_n$. The product
$\otimes_M$ is well defined on sufficiently smooth functions and
extends continuously to the Schwartz space. In terms of the
Fourier transforms, the relation between the Moyal tensor product
and the ordinary one takes the simple form
\begin{equation*}
(\widehat{f\otimes_M g})(k,k') = e^{-\tfrac{i}{2}k\theta k'} (\hat
f\otimes \hat g)(k,k'), \quad f,g\in S(\oR^d).
\end{equation*}
The exponential factor here is a multiplier for $S(\oR^{2d})$ and
hence defines an automorphism of this space. An analogous relation
holds for $f_1\otimes_M\dots\otimes_M f_n$ with the multiplier
\begin{equation}
\mu_n(k)= \prod_{1\leq a<b\leq n} e^{-\tfrac{i}{2}k_a\theta k_b}.
 \label{10}
\end{equation}
This together with the Schwartz kernel theorem implies that for
every continuous multilinear  form
\begin{equation*}
v\colon \underbrace{S(\oR^d)\times\dots\times
S(\oR^d)}_n\rightarrow \mathbb{C},
\end{equation*}
there is a unique linear functional $w\in S'(\oR^{nd})$ such that
the diagram
 \begin{equation}
\xymatrix{
 S(\oR^d)\times\dots\times S(\oR^d)\ar[r]^(0.75)v\ar[d]_{\otimes_M}&\oC\\
 S(\oR^{nd})\ar[ur]_w
 }
 \label{11}
  \end{equation}
is commutative.

\section{Twist-deformed quantum fields}
\label{sec3}

We now turn to the noncommutative deformation~\cite{8,9}  of
quantum field theories that can be associated with the Moyal
tensor product. This construction  does not use the Lagrangian
formulation  and is model independent, but  we here restrict our
consideration to the  case of one scalar field $\phi(x)$ for
simplicity. Our starting point is a quantum field theory on
commutative Minkowski space with the usual
assumptions~\cite{17,18} of relativistic covariance, locality, and
physical stability expressed by the spectral condition. We also
make the standard assumptions about the domain and continuity of
the field. Then every vacuum expectation value  is well defined as
a tempered distribution
\begin{equation}
\langle \Psi_0,
\phi(f_1)\cdots\phi(f_n)\Psi_0\rangle=(w^{(n)},f_1\otimes\cdots\otimes
f_n), \qquad w^{(n)}\in S'({\mathbb R}^{4n}).
 \label{12}
\end{equation}
The idea is to  deform these distributions  by replacing the
tensor product of their arguments with the twisted tensor product
\begin{equation}
(w_\theta^{(n)},f_1\otimes\cdots\otimes
f_n)\eqdef(w^{(n)},f_1\otimes_M\cdots\otimes_M f_n).
 \label{13}
\end{equation}
It follows from what has been said above that there exists a
unique $w_\theta^{(n)}\in S'(\oR^{4n})$ satisfying~\eqref{13} and
by the Wightman reconstruction theorem, the sequence of deformed
distributions $w_\theta^{(n)}$ can be used to construct a field
theory. It is notable that there is  in fact no need to appeal to
the reconstruction theorem because we can easily directly define a
quantum field $\phi_\theta$ whose vacuum expectation values
coincide with the deformed Wightman functions $w_\theta^{(n)}$.
Namely,  the Schwartz kernel theorem gives a precise meaning to
all vectors of the form
\begin{equation}
\Phi_n (f)=\int dx_1\dots dx_n\, \phi(x_1)\cdots\phi(x_n)f(x_1,
\dots, x_n)\Psi_0,
\label{14}
\end{equation}
where $\Psi_0$ is the vacuum state and $f$ ranges  the space
$S({\mathbb R}^{dn})$. Therefore, we may take the linear span $D$
of all such vectors as a domain  of the initial field $\phi$. For
each $g\in S(\oR^4)$, we define $\phi_\theta(g)$ by
\begin{equation}
\phi_\theta(g)\Psi_0=\phi(g)\Psi_0,\qquad
\phi_\theta(g)\Phi_n(f)=\Phi_{n+1}(g\otimes_M f),\quad n\geq
1,\label{15}
\end{equation}
extended by linearity. The properties of the fields $\phi_\theta$
can be summarized as follows~\cite{8}.

{\bf Theorem~1.} \label{th1} {\it Let $\phi$ be a Hermitian scalar
field satisfying the Wightman axioms and let $w^{(n)}$ be its
Wightman functions. Then the deformed fields $\phi_\theta$ are
well defined as operator-valued tempered distributions with the
same domain in the Hilbert space of $\phi$,  and
\begin{equation}
\langle
\Psi_0,\,\phi_\theta(g_1)\cdots\phi_\theta(g_n)\Psi_0\rangle=
(w^{(n)}, \,g_1\otimes_M\cdots\otimes_M g_n),\qquad g_j\in
S(\oR^4). \label{16}
 \end{equation}
The vacuum state $\Psi_0$ of $\phi$ is a cyclic vector for every
field $\phi_\theta$. These fields satisfy the hermiticity
condition
\begin{equation}\phi_\theta(g)^*\supset\phi_\theta(\bar g),\qquad
g\in S({\mathbb R}^4),
\label{17}
\end{equation}
and their vacuum expectation values $w_\theta^{(n)}$  satisfy the
spectral condition as well as the positive definiteness condition
\begin{equation}
\sum\limits_{m,n=0}^N (w^{(m+n)}_\theta,\,f^*_m\otimes f_n)\geq 0,
\label{18}
\end{equation}
where  $w^{(0)}_\theta=1$, $f_0\in \oC$,  $f^*(x_1,\dots,
x_m)\eqdef \overline{f(x_m,\dots, x_1)}$,  and $\{f_j\}_{1\leq
j\leq N}$ is an arbitrary finite set of test functions such that
$f_j\in S(\oR^{4j})$.

Proof.} For any  $g\in S(\oR^4)$, $f\in S(\oR^{4n})$, and $h\in
S(\oR^{4m})$, we have
\begin{equation}
\langle\Phi_m(h),\Phi_{n+1}(g\otimes_M f)\rangle=
\langle\Phi_{m+1}(\bar g\otimes_M h),\Phi_n(f)\rangle,
 \label{19}
 \end{equation}
 or, equivalently,
\begin{equation}
(w^{(m+n+1)},\,h^*\otimes(g\otimes_M f))=(w^{(m+n+1)},\,(\bar
g\otimes_M h)^*\otimes f).
 \label{20}
 \end{equation}
Indeed, it follows  from  definition~\eqref{9} that $(\bar
g\otimes_M h)^*=h^*\otimes_M g$  because the matrix  $\theta$ is
antisymmetric. Furthermore,
\begin{gather*}
\hat h^*\otimes(\widehat{g\otimes_M f})=(\hat h^*\otimes \hat
g\otimes \hat f)e^{-\tfrac{i}{2}k\theta(\sum_{a=1}^np_a)},\\
(\widehat{h^*\otimes_M g})\otimes \hat f=(\hat h^*\otimes \hat
g\otimes \hat f)e^{-\tfrac{i}{2}(\sum_{b=1}^mq_b)\theta k},
\end{gather*}
where $k$, $p_a$, and $q_b$ are the respective  arguments of the
functions $\hat g$, $\hat f$, and   $\hat h^*$. Again using the
antisymmetry of $\theta$, we obtain relation~\eqref{20} because
$\hat w^{(m+n+1)}(q_1,\dots, q_m, k, p_1,\dots,p_n)$ contains the
factor $\delta(k+\sum_{b=1}^m q_b+\sum_{a=1}^n p_a)$ expressing
the translation invariance of $w^{(m+n+1)}$. It follows
from~\eqref{19} that if a linear combination $\Phi$ of vectors of
form~\eqref{14} is zero, then $\phi_\theta(g)\Phi$ is also zero,
i.e., the operators $\phi_\theta(g)$ are well defined. It is also
obvious that $\phi_\theta(g)\Phi\to 0$ as $g\to 0$ in $S(\oR^4)$.
Moreover, \eqref{19} implies~\eqref{17}.

Identity~\eqref{16} follows directly from~\eqref{15}. If a vector
$\Psi$ is such that
$\langle\Psi,\,\phi_\theta(g_1)\cdots\phi_\theta(g_n)\Psi_0\rangle=0$
for arbitrary $g_1,\dots, g_n\in S(\oR^4)$ and each $n$, then
$\langle\Psi,\,\Phi_n(f)\rangle=0$ for every $\Phi_n(f)$ of
form~\eqref{14} because $\langle\Psi,\,\Phi_n(f)\rangle$ is just
the element of  $S'(\oR^{4n})$ that is associated with the
multilinear form $(g_1,\dots,g_n)\to
\langle\Psi,\,\phi_\theta(g_1)\cdots\phi_\theta(g_n)\Psi_0\rangle$
by  diagram~\eqref{11}. Therefore, $\Psi=0$, and  the linear span
of vectors $\phi_\theta(g_1)\cdots\phi_\theta(g_n)\Psi_0$ is hence
dense in the Hilbert space.  Clearly,  $\hat
w^{(n)}_\theta=\mu_n\cdot\hat w^{(n)}$, where  $\mu_n$ is given
by~\eqref{10}, and  the deformed Wightman functions therefore have
the same spectral properties as those of $w^{(n)}$'s. To
prove~\eqref{18}, we note that $(w^{(m+n)}_\theta,\,f^*_m\otimes
f_n)=(w^{(m+n)},\,g_m^*\otimes_M g_n)$, where $\hat
g_m=\mu_m\cdot\hat f_m$ and $\hat g_n=\mu_n\cdot\hat f_n$.
Furthermore, by  the antisymmetry of $\theta$ and the translation
invariance of $w^{(m+n)}$, we have $(w^{(m+n)},\,h\otimes_M
g)=(w^{(m+n)},\,h\otimes g)$ for each $h\in S(\oR^{dm})$ and $g\in
S(\oR^{dn})$ because $\widehat{h\otimes_M g}= (\hat h\otimes \hat
g)e^{-\tfrac{i}{2}(\sum_1^mq_b)\theta(\sum_1^np_a)}$. The theorem
is proved.

By the standard argument~\cite{17}, any other implementation of
field theory with  vacuum expectation values~\eqref{16} and with a
cyclic vacuum state invariant under translations is unitary
equivalent to this one. If the initial field $\phi$ is free, than
its creation and annihilation operators  are deformed as follows
\begin{equation*}
a_\theta(p)=e^{\tfrac{i}{2}p\,\theta P} a(p),\qquad
a^*_\theta(p)=e^{-\tfrac{i}{2}p\,\theta P} a^*(p),
\end{equation*}
where  $P$ is the energy-momentum operator. Accordingly, they
satisfy the deformed canonical commutation relations (CCRs)
\begin{gather}
a_\theta(p)a_\theta(p')=e^{-ip\theta
p'}a_\theta(p')a_\theta(p),\quad
a^*_\theta(p)a^*_\theta(p')=e^{-ip\theta
p'}a^*_\theta(p')a^*_\theta(p), \notag
\\
a_\theta(p)a^*_\theta(p')=e^{ip\theta p'}a^*_\theta(p')
a_\theta(p)+2\omega_{\bf p}\delta(\bf p- \bf p').
 \label{21}
 \end{gather}
Twisted CCR algebra~\eqref{21} was discussed from different
standpoints in~\cite{6,19,20,21,22}. As already stated, this
deformation preserves the translation invariance, but it obviously
violates the Lorentz invariance. The full transformation law of
the deformed fields under the proper Poincar\'e group is given by
\begin{equation}
U(a,\Lambda)\phi_\theta(f)U^{-1}(a,\Lambda)=
\phi_{\Lambda\theta\Lambda^T}(f_{a,\Lambda}),
 \label{22}
 \end{equation}
where $f_{(a,\Lambda)}(x)= f(\Lambda^{-1}(x-a))$, $(a,\Lambda)\in
\mathcal P_+^\uparrow$. Hence, every  $\phi_\theta$ transforms
covariantly only under those Lorentz transformations that leave
the matrix $\theta^{\mu\nu}$ unchanged.

 This deformation also leads  to
the lack of locality and the fields $\phi_\theta$ do not satisfy
the standard microcausality condition. This can be easily seen by
considering the matrix elements
\begin{equation}
M_\Phi(x,x')=\langle \Psi_0, [\phi_\theta(x),\phi_\theta(x')]\,
 \Phi\rangle
 \label{23}
 \end{equation} in
the simplest case of a free field. If we take $\Phi$ to be a
normalized  two-particle state
\begin{equation}
\Phi=
 \phi^{(-)}(h)\phi^{(-)}(h)\,\Psi_0,\qquad\mbox{where}\quad
 h\in S({\mathbb R}^d),
 \label{24}
  \end{equation}
then a simple direct calculation shows that the matrix
element~\eqref{23} is nonzero for some spacelike separated points.
Moreover, as shown in~\cite{8}, this distribution does not vanish
in any open region, i.e., $\supp M_\Phi={\mathbb R}^{2\cdot4}$.
The reason is that its Fourier transform  has the form
\begin{equation*}
\widehat{M}_\Phi(k,k')=-4 i \hat w(k)\hat w(k') \hat h(k) \hat
h(k')\sin\frac{k\theta k'}{2},
\end{equation*}
where $\hat w(k)$ is the momentum-space two-point function of
$\phi$. Hence ${\widehat M}_\Phi$ has support in the product $\bar
\oV_+\times\bar\oV_+$ of two closed forward cones. The factor
$(\hat h\otimes \hat h)\sin\dfrac{k\theta k'}{2}$ does not vanish
on $\supp (\hat w\otimes\hat w)$ if $\Phi\ne 0$. Therefore,
$M_\Phi$ is the boundary value of a nonzero analytic function.
Such a distribution cannot vanish on any open nonempty set by the
general uniqueness theorem for analytic functions (see Theorem~B.7
in~\cite{18}). We can therefore conclude that  the noncommutative
deformation by twisting products in the vacuum expectation values
brings the causality principle and the spectral condition into
conflict.

\section{Localization in wedge-shaped regions}
\label{sec4}

The deformed fields $\phi_\theta$ with different $\theta$ retain
some relative localization properties, established in~\cite{9,21},
where it was noted that a wedge-shaped space-time region
$W_\theta$ can be associated with each antisymmetric matrix
$\theta$. If this matrix has the standard form
\begin{equation*}
\theta_1=\begin{pmatrix}
   0&\vartheta_e & 0& 0\\
-\vartheta_e & 0& 0& 0\\
0&0&0&\vartheta_m\\
0&0&-\vartheta_m &0
\end{pmatrix},\qquad \vartheta_e\ge0,
\end{equation*}
then its associated wedge $W_1$ is defined by the inequality
$x^1>|x^0|$. The stabilizer subgroup of $\theta_1$ with respect to
the action $\theta\to\Lambda\theta\Lambda^T$ of the proper Lorentz
group $\mathcal L_+$ coincides with that  of $W_1$ with respect to
the action $W\to\Lambda W$, and  there is therefore a one-to-one
correspondence between the orbits of $\theta_1$ and $W_1$. Grosse
and Lechner~\cite{9,21} have shown that the  fields $\phi_\theta$
satisfy the following wedge-local commutation relation: if the
sets $\supp f+W_\theta$ and $\supp g+W_{\theta'}$ are spacelike
separated, then
\begin{equation*}
[\phi_\theta(f), \phi_{\theta'}(g)]\,\Psi=0\qquad \mbox{for
all}\quad \Psi\in D.
\end{equation*}
Therefore, the deformed fields should be regarded as objects
localizable   in wedge-shaped regions and not at points of
space-time.

Such a localization   is similar to that studied in~\cite{23} in
the framework of algebraic quantum field theory. As noted
in~\cite{9,21}, even this weak form of local commutativity allows
constructing a scattering theory. But this type of localization is
radically different from the replacement of the light cone with a
light wedge proposed previously as a possible modification of the
microcausality condition in field theories with space-space
noncommutativity corresponding to the low-energy limit of string
theory.  In the next section, we discuss an instructive example
giving an idea of such a modification.

\section{Twist-deformed  Wick square of a free field}
\label{sec5}

We consider the normal ordered square of a free scalar field in
the four-dimensional space-time but change the ordinary product in
its definition to the Moyal star product:
\begin{multline}
\mathcal O(x)\stackrel{\text{\tiny def}}{=}\,:\phi\star_M\phi:(x)
=\lim_{x_1,x_2\to
x}:\phi(x_1)\phi(x_2):\\+\sum_{n=1}^\infty\left(\frac{i}{2}
\right)^n\frac{1}{n!}\,\theta^{\mu_1\nu_1}\dots
\theta^{\mu_n\nu_n}\lim_{x_1,x_2\to x}:\partial_{\mu_1}\dots
\partial_{\mu_n}\phi(x_1)\,\partial_{\nu_1}\dots\partial_{\nu_n}\phi(x_2):.
 \label{25}
\end{multline}
It is easy to see that in the case of space-space
noncommutativity, where $\theta^{23}=-\theta^{32}=\vartheta\ne 0$
and the other elements of the  matrix $\theta$ are  zero, the
commutator $[\mathcal O(x), \mathcal O(y)]$ vanishes in the wedge
defined by the inequality $|x^0-y^0|<|x^1-y^1|$. It was shown
in~\cite{24} that an analogous commutator with the time derivative
is nonzero at some points outside this wedge:
\begin{equation*}
\langle 0|\,[\mathcal O(x), \partial_0\mathcal
O(y)]_-|p_1,p_2\rangle|_{x^0=y^0}\ne 0.
\end{equation*}
It was conjectured in that paper that this result holds generally
when there are time derivatives in the observables. A closer
examination in~\cite{8} showed that the space-space
noncommutativity violates the microcausality condition even if
there are no time derivatives in the observables. Moreover, the
commutator $[\mathcal O(x),\mathcal O(y)]$  is nonzero for any
points $\bar x$ and $\bar y$  outside the wedge
$|x^0-y^0|<|x^1-y^1|$. In more precise terms, there is a state
$\Phi$ such that $(\bar x,\bar y)$ belongs to the support of the
distribution $\langle \Psi_0,[\mathcal O(x),\mathcal
O(y)]\,\Phi\rangle$.

The star commutator
\begin{equation*}
[\mathcal O(x),\mathcal O(y)]_\star= \mathcal O(x)\star_M\mathcal
O(y)-\mathcal O(y)\star_M\mathcal O(x)
\end{equation*}
was also discussed in the literature with contradictory
conclusions. An analysis of its matrix elements in the same
paper~\cite{8} showed that it is nonzero everywhere. In
particular, these results demonstrate that  the power series
expansions of the matrix elements  in the noncommutativity
parameter do not converge in the topology of the space of tempered
distributions, because their partial sums obviously satisfy the
microcausality condition.

This simple example also  demonstrates that noncommutative field
theory with the Wick-Voros star product is more nonlocal than that
with the Moyal product.  A calculation analogous to that
in~\cite{8} shows that in the Wick-Voros case, the matrix element
of the corresponding commutator between the vacuum and
two-particle state~\eqref{24} has the form
\begin{multline}
\langle \Psi_0,[\mathcal O_V(x),\mathcal O_V(y)]\,\Phi\rangle=
8\!\iiint\!\!\epsilon
(k^0)\delta(k^2-m^2)e^{\tfrac{\vartheta}{2}{\bf k(p_2-p_1)}}
e^{-ik\cdot (x-y)-ip_1\cdot x-ip_2\cdot
y}\\
\times\prod_{a=1}^2\theta(p_a^0)\delta(p_a^2-m^2)
\cos\left(\frac{1}{2}k\theta  p_a\right)\hat h(p_a)
\frac{dkdp_1dp_2}{(2\pi)^{12}},
 \notag
\end{multline}
where ${\bf k}=(k^2, k^3)$. This expression contains an additional
exponential factor $e^{\tfrac{\vartheta}{2}{\bf k(p_2-p_1)}}$
compared with the Moyal case and is therefore not a tempered
distribution and is well defined in coordinate space only on
analytic test functions.

\section{Convergence of  star products and  adequate spaces of \\functions}
\label{sec6}

We now turn to the  conditions for the  convergence of the star
products and to the description of the test function
spaces~\cite{13,14} that are most suitable for a noncommutative
quantum field theory. As stated above, the formal power series
representing the Moyal and Wick-Voros star products  generally
diverge for  functions belonging to the Schwartz space. But both
of them converge under the additional condition
\begin{equation}
(1+|x|)^N |\partial^\kappa f(x)|<C_N B^{|\kappa|}\sqrt{\kappa!},
\label{26}
\end{equation}
where $C_N$ and $B$ are constants depending on $f$ and
\begin{equation}
B<\frac{1}{\sqrt{|\theta|}},\qquad |\theta|=\max|\theta^{\mu\nu}|.
\label{27}
\end{equation}
In~\eqref{26},  the  multi-index notation
$\kappa=(\kappa_1,\dots,\kappa_d)\in \oZ_+^d$,  $\partial^\kappa
=\partial^{\kappa_1}_{x_1}\dots\partial^{\kappa_d}_{x_d}$,
$|\kappa|=\kappa_1+\dots+\kappa_d$, and
$\kappa!=\kappa_1!\dots\kappa_d!$ is used. The convergence of the
Moyal product under the indicated condition was proved in two
different ways in~\cite{13,14}. But the function space defined
by~\eqref{26} is not an algebra   even with respect to the usual
pointwise multiplication. To obtain an algebra, we take those
functions that satisfy analogous inequalities with an arbitrary
small $B$ and let  $\mS^{1/2}$ denote this space. It has a natural
topology determined by the set of norms
\begin{equation*}
\|f \|_{B,N}=\sup_{x,\kappa}\,(1+|x|)^N\frac{|\partial^\kappa
f(x)|}{B^{|\kappa|}\sqrt{\kappa!}}.
\end{equation*}
It is easy to see that $\mS^{1/2}$ is a Fr\'echet space, i.e., is
metrizable and complete. Moreover, it is a nuclear space, as
follows from a  result in~\cite{25}, where it was established that
the Gelfand-Shilov spaces $S^\beta$ are nuclear. But we emphasize
that $\mS^{1/2}\ne S^{1/2}$. The space $S^{1/2}$ is the inductive
limit of the family of spaces $S^{1/2, B}$ as $B\to \infty$ (and,
in particular, is nonmetrizable), while $\mS^{1/2}$ is the
projective limit of the same family as $B\to 0$. The spaces
$S^\beta$ with $\beta<1/2$ were  proposed for using in
noncommutative QFT in~\cite{26}. The nuclearity  is crucial in
deriving the Schwartz kernel theorem, which plays an important
role in axiomatic formulation of quantum field theory. A simple
proof of an analogous kernel theorem for a large class of spaces
including $\mS^{1/2}$ and $S^\beta$ was given in~\cite{27,28}. The
following result extends Theorem~6 in~\cite{14}.

{\bf Theorem~2.} {\it The space $\mS^{1/2}({\mathbb R}^d)$ is a
topological algebra with respect to both the Moyal star product
and  the Wick-Voros star product. If $f,g \in \mS^{1/2}({\mathbb
R}^d)$, then the series representing these products   converge
absolutely in this space. Moreover these products depend
continuously on the noncommutativity parameter $\theta$, and the
operator $T= e^{\tfrac{\vartheta}{4}\Delta}$ is an automorphism of
this space.

Proof.} We first describe the infinite-order differential
operators
\begin{equation}
\sum_{\lambda\in \oZ_+^d} c_\lambda \partial^\lambda
 \label{28}
\end{equation}
that act continuously on $\mS^{1/2}$. We assume that $\sum_\lambda
c_\lambda z^\lambda$ has less than exponential growth of order $2$
and type $b$, which is equivalent to the condition
\begin{equation}
|c_\lambda|\le C(2b)^{|\lambda|/2}\frac{1}{\sqrt{\lambda!}}.
\label{29}
\end{equation}
Applying \eqref{28} to $f\in \mS^{1/2}$,  we obtain
\begin{multline}
(1+|x|)^N\left|\partial^\kappa\sum_\lambda
c_\lambda\partial^\lambda f(x)\right|\le \|f \|_{B,N}\sum_\lambda
c_\lambda B^{|\kappa+\lambda|}\sqrt{(\kappa+\lambda)!} \\
\leq \|f
\|_{B,N}2^{|\kappa|/2}B^{|\kappa|}\sqrt{\kappa!}\sum_\lambda
c_\lambda 2^{|\lambda|/2}B^{|\lambda|}\sqrt{\lambda!}.
 \label{30}
\end{multline}
We also assume that
\begin{equation}
b<\frac{1}{4B^2}.
 \label{31}
\end{equation}
Then the last series in~\eqref{30}  converges, and taking  $B'\ge
 B\sqrt2$, we obtain
 \begin{equation*}
 \left\|\sum_\lambda
c_\lambda\partial^\lambda f\right\|_{B',N}\le C'\|f \|_{B,N}.
\end{equation*}
Therefore,   operator~\eqref{28}  with any  $b<\infty$ maps
$\mS^{1/2}$ into itself continuously. We note that  the operator
in definition~\eqref{2} of the $\star_M$-product has a type of
growth that is obviously no greater than $|\theta|/4$
and~\eqref{27} hence implies \eqref{31}. If the matrix $\theta$
has the standard form indicated  in Sec.~II, then $b$ is simply
 $\vartheta$. We conclude that $T=
e^{\vartheta\Delta/4}$ is an automorphism of $\mS^{1/2}(\oR^d)$.

Further, the operators  in definitions~\eqref{7} and \eqref{8} are
automorphisms of $\mS^{1/2}(\oR^{nd})$, and the series
representing the twisted tensor products converge absolutely in
$\mS^{1/2}(\oR^{nd})$. Using the Leibnitz formula, we can easily
verify that $\mS^{1/2}$ is a topological algebra under the
pointwise multiplication. It follows that the  map
$\mS^{1/2}(\oR^{2d})\to \mS^{1/2}(\oR^d)\colon f(x,y)\to f(x,x)$
is continuous because this map is just the linear  map associated
with the pointwise multiplication and
$\mS^{1/2}(\oR^{2d})=\mS^{1/2}(\oR^d)\hat{\otimes}\mS^{1/2}(\oR^d)$
by the kernel theorem. The Moyal and Wick-Voros star
products~\eqref{2} and \eqref{3} are obtained from $f\otimes_Mg$
and $f\otimes_Vg$ by this map and   are therefore also continuous.
Their  continuity in $\theta$ can be established in the same way
as in the proof of Theorem~5 in~\cite{13}. The theorem is proved.

It is essential that $\mS^{1/2}$ is  the largest  subspaces of the
Schwartz space $S$ with the properties established in Theorem~2.
But using only $\mS^{1/2}$ does not suffice to  describe the
nonlocal features introduced by noncommutativity exactly. To
better understand the causal structure of noncommutative models,
it is useful to examine the decay properties of the field
commutators more closely. A simple, well-known way to describe the
decay properties of a distribution is to consider its convolution
with sufficiently rapidly decreasing test functions. But the
functions satisfying the  convergence condition stated above
cannot decrease too rapidly. The most suitable space consists of
functions with a Gaussian decreas. More precisely, it is
reasonable to use a two-parameter family of spaces $S^{1/2,
B}_{1/2,A}$, each consisting of functions with the finite norm
\begin{equation*}
\|f\|_{A,B}=\sup_{\kappa, x}e^{|x/A|^2}\frac{|\partial^\kappa
f(x)|}  {B^{|\kappa|} \sqrt{\kappa!}}
\end{equation*}
As shown in the appendix in~\cite{8}, this space is nontrivial if
$AB>2$ and becomes trivial if  $AB<\sqrt2$.

\section{$\theta$-locality}
\label{sec7} Using rapidly decreasing  test functions, we can show
that certain causality properties hold independently of the type
of noncommutativity. For this, we again consider the deformed
normal ordered square ${\mathcal O}(x)=:\phi\star\phi:(x)$ of a
free field but  average it with such functions this time,
\begin{equation*}
\mathcal O(f_a)=\int dx \,\mathcal O(x)f(x-a),\qquad f\in S^{1/2,
B}_{1/2,A}.
\end{equation*}
We must estimate the asymptotic behavior of the commutator
$[\mathcal O(f_a),\mathcal O(g_{-a})]$ of two averaged observables
at large spacelike separations. We fix a spacelike vector $a$ and
let $\gamma$ denote  the angular distance of this vector from the
light cone,
\begin{equation*}
\gamma=\inf_{\xi^2\geq 0}|\xi-a/|a||.
\end{equation*}
Evaluating the matrix element
\begin{equation}
\langle \Psi_0,[\mathcal O(f_a),\mathcal O(g_{-a})]\,\Phi\rangle
 \label{32}
\end{equation}
between the vacuum and a two-particle state
$\Phi=\phi^{(-)}(h_1)\phi^{(-)}(h_2)\Psi_0$, we find~\cite{8} that
it decreases  in a Gaussian manner as  $|a|$ increases:
\begin{equation}
  |\langle \Psi_0,[\mathcal O(f_a),\mathcal O(g_{-a})]\,\Phi\rangle|\leq C_{\Phi,A'}\|f\|_{A,B}\|g\|_{A,B}
   e^{-2|\gamma\, a/A'|^2}
 \label{33}
\end{equation}
for all $A'>A$. It follows from  the convergence condition
$B<1/\sqrt{|\theta|}$ and the nontriviality condition $AB>2$, that
the best result is at $A\sim\sqrt{|\theta|}$. We therefore
conclude that  matrix element~\eqref{32} decreases like
$e^{-|\gamma a|^2/|\theta|}$ as the spacelike separation  tends to
infinity. This  result holds for both  the Moyal and the
Wick-Voros  products. Moreover, it can be shown that any matrix
element of the commutator behaves similarly.

Estimate~\eqref{33} is quite informative, but it can be expressed
in more abstract terms that are even more convenient for
applications. This can be done   using another class of function
spaces which are associated with cone-shaped regions and defined
as follows. Let $U$ be a cone in $\oR^d$. A smooth function $f$ on
${\mathbb R}^d$ belongs to the space $S^{1/2,B}(U)$  if it
satisfies the condition
\begin{equation*}
\sup_{x\in U}(1+|x|)^N |\partial^\kappa f(x)|<C_N
B^{|\kappa|}\sqrt{\kappa!}.
\end{equation*}
It turns out that the result~\eqref{33} is equivalent to saying
that for any $B<1/\sqrt{|\theta|}$, the matrix element considered
as a generalized function has a continuous extension to the space
$S^{1/2,B}(\mathbb{V})$ associated with the relative light cone
$\mathbb{V}=\{(x,y)\in {\mathbb R}^4\times {\mathbb R}^4\colon
(x-y)^2\geq 0\}$. This suggests how the microcausality condition
could be generalized to quantum fields on noncommutative
space-time.

The nonlocal effects in noncommutative field theory are determined
by the structure of the  star product, and we can expect that in
this theory, all matrix elements
$\langle\Phi\,,[\phi(x),\psi(x')]_-\Psi\rangle$ of field
commutators (or anticomutators for unobservable fields) allow a
continuous extension to such a space $S^{1/2,B}(\mathbb{V})$ whose
superscript $B$ is of the order $1/\sqrt{|\theta|}$ but in general
may depend on the fields $\phi$ and $\psi$ and the states $\Phi$
and $\Psi$. This condition has been introduced in~\cite{14} and we
call it $\theta$-{\it locality} for brevity. We emphasize that it
is consistent with the Poincar\'e covariance. Conceivably,
$\theta$-locality expresses the absence of acausal effects on
scales much larger than the fundamental length $\sqrt{|\theta|}$.
If such is indeed the case, then this formulated condition might
be called macrocausality. It is quite possible that  the matrix
elements of fields are tempered distributions in physically
relevant noncommutative models. Even so, the space
$S^{1/2,B}(\oV)$ can be considered as a tool for formulating
causality, and  it is clear from the above that its role can be
more essential in the case of the Wick-Voros star product.

\section{Conclusion}
\label{sec8} The physical consequences of the noncommutative
quantum field theory obtained by  twisting  tensor products need
further investigation. Here, we did not touch the twisted
Poincar\'e covariance, but we note that  naively coupling  the
deformation considered above with this covariance can lead to a
theory  physically equivalent to the undeformed one, analogously
to the theory discussed in~\cite{20}. The issue of how best to
combine the idea of space-time noncommutativity with the (properly
adapted) basic principles of quantum physics  still  remains open.
The  $\theta$-locality condition   means  that the commutators of
observables behave at large spacelike separations like
$\exp(-|x-y|^2/\theta)$, and this condition is similar to the
asymptotic commutativity condition  previously used in nonlocal
QFT on the usual commutative Minkowski space. As shown
in~\cite{29}, the asymptotic commutativity in combination with the
relativistic covariance and the spectral condition ensures the
existence of the CPT symmetry and the usual spin-statistics
relation for nonlocal fields. This is an additional  argument for
using the $\theta$-locality to formulate causality in a
noncommutative quantum field theory. The space $\mS^{1/2}$,  being
a maximal topological star product algebra with absolute
convergence,   is completely adequate to the concept of strict
deformation quantization and can be used to formulate and study
noncommutative field models nonperturbatively.

\subsection*{Acknowledgements}
The author thanks the Organizing Committee and Prof.~I.~Aref'eva
for the kind invitation to give a talk at  the Second
International Conference "String Field Theory and Related Aspects"
(Moscow, 2009).

 This work was supported in part by the the Russian
Foundation for Basic Research (Grant No.~09-01-00835) and the
Program for Supporting Leading Scientific Schools (Grant
No.~NSh-1615.2008).

\end{document}